\documentclass[preprint,amsmath,amssymb,amsthm,aps]{revtex4-2}
\usepackage{graphicx}
\usepackage{capt-of}
\usepackage{lipsum}
\usepackage{dcolumn}
\usepackage{bm}
\usepackage{hyperref}
\usepackage{float}
\usepackage{xcolor}
\usepackage{color}
\usepackage{placeins}
\usepackage{subeqnarray}
\usepackage{empheq}
\usepackage{ntheorem}
\theoremseparator{:}

\usepackage{cases}
\usepackage{lineno}
\usepackage{fancyhdr}
\begin{document}

\title{Dynamics of motions and deformations of an arbitrary geometry flexural floe in ocean waves}
\author{A. Ludu}
\email{ludua@erau.edu}
\affiliation{Embry-Riddle Aeronautical University, Dept. of Mathematics \& Wave Lab\\
	Daytona Beach, FL 32124 USA}

\date{\today}

\begin{abstract}
This paper develops a comprehensive mathematical framework for modeling the coupled hydroelastic dynamics of sea-ice floes of arbitrary shape and non-uniform thickness under linear ocean wave forcing. We simultaneously incorporate four dominant rigid-body motions (heave, surge, roll, pitch) and the complete spectrum of flexural deformation modes within a unified Green function formulation. The water flow is modeled using potential theory with Laplace's equation, while the floe obeys a generalized Kirchhoff-Love plate equation with spatially varying flexural rigidity. We formulate the coupled fluid-structure interaction problem through kinematic velocity-matching conditions and dynamic pressure-continuity conditions at the ice-water interface. The elastic eigenproblem with free-edge boundary conditions yields a complete orthogonal basis of deformation modes, accounting for added mass effects through modified natural frequencies. By decomposing the velocity potential into partial potentials associated with incident waves, scattered waves, rigid motions, and elastic modes, we reduce the problem to a system of Fredholm integral equations of the second kind for surface density functions on all boundary segments. The solution methodology employs single-layer potential representations with fundamental Green functions for Laplace's equation. We present explicit formulations for all boundary conditions in compact tensor form, provide asymptotic analysis for the spectrum of non-uniform thickness floes, and discuss resonance phenomena arising from the interaction between incident wave frequency and natural vibration modes.
\end{abstract}


\maketitle

\section{Introduction}
\label{sec1}

The interaction between ocean waves and sea ice represents a fundamental problem in
polar ocean dynamics, with significant implications for climate modeling, maritime operations,
and understanding the rapidly changing Arctic and Antarctic environments. This
report provides a comprehensive review of the mathematical-physical theoretical framework
for modeling the dynamics of sea-ice floes under linear plane ocean waves, with
particular emphasis on Green function methods. Sea-ice floes in the marginal ice zone exhibit complex hydroelastic behavior when subjected to ocean wave forcing. The floes simultaneously undergo rigid-body motions
(heave, surge, sway, roll, pitch, and yaw) and elastic deformations that can lead to
fracture. Understanding these coupled dynamics is essential for predicting wave attenuation
in ice-covered seas, ice breakup patterns, and the evolution of floe size distributions.

This topic  attracted much attention in the recent years. In water wave theory the most studied situation, both by analytical and numerical techniques, is the wave interaction with a single or with an array  of floating of fixed bodies with axial symmetry. In the complex dynamics of floating systems, the main parameters to take into account are the nature of the wave field, and the geometry and material properties of the body.  Clearly, in order to obtain a more realistic model one can work in a larger parameter spaces, including currents, bathymetry, wind interaction, thermodynamics of the sea, wave run-up during unavoidable over-topping events in rough seas, etc.

The mathematical modeling of sea-ice floe dynamics under ocean wave forcing has evolved significantly over the past several decades, driven by advances in both computational methods and analytical techniques for fluid-structure interaction problems. The foundational work on floating body dynamics in the naval architecture community, particularly the seminal contributions in~\cite{john1, john2} and~\cite{fenton,isaacson}, established the theoretical framework for analyzing rigid body motions of axisymmetric structures in waves. These early studies focused primarily on six degrees of freedom for rigid body motion---heave, surge, sway, roll, pitch, and yaw---treating the floating body as completely rigid and often restricting attention to bodies with circular horizontal cross-sections.

The extension of these methods to ice floe modeling began with investigations of wave attenuation in ice-covered seas, where researchers recognized that elastic deformation of the floe could play a crucial role. In~\cite{massonleblond} the authors studied spectral evolution of waves in dispersed ice fields, while 
in~\cite{squire} the authors provided a comprehensive review of wave-ice interaction phenomena, highlighting the hydroelastic nature of the problem. The Green function method emerged as a particularly powerful approach for these problems, as demonstrated by a series of influential papers~\cite{meylan1996a,meylan2002,meylan2015bc,meylan2021,meylan2021good,x1,x2,x3}. These works treated circular ice floes with uniform thickness, solving coupled problems involving rigid body motions and elastic flexural deformations using potential flow theory for the water and thin plate theory for the ice.

The treatment of elastic deformations in ice floes has traditionally employed either the classical Kirchhoff-Love plate theory or its generalizations. For uniform thickness plates, the eigenvalue problem associated with free edge boundary conditions admits clean separation of variables in polar coordinates, yielding orthogonal basis functions characterized by radial and circumferential mode numbers~\cite{lei,lu,la}. Several studies have computed natural frequencies and mode shapes for circular plates, both in vacuum and in contact with fluids~\cite{books,ew1,ew2,ew3,ew4,ew5,ew6,flop,evans,yeung,senjan}, providing essential data for understanding added mass effects and fluid-structure coupling.

From a hydrodynamic perspective, the standard approach follows classical potential flow theory with appropriate boundary conditions on the free surface, seabed, and fluid-structure interface. The linearized Bernoulli equation relates pressure to the velocity potential, while kinematic boundary conditions ensure velocity continuity at interfaces. For problems involving scattered waves, absorbing boundary conditions at artificial far-field boundaries are essential; these range from simple Sommerfeld conditions~\cite{bc1,bc2,dtn,somm1,hist} to more sophisticated approaches such as Bayliss-Turkel or Engquist-Majda conditions~\cite{bc1,bc2,dtn}, and exact Dirichlet-to-Neumann boundary conditions~\cite{dtn}.

The solution methodology typically employs either boundary integral equations or Green function representations. The Green function approach, utilizing fundamental solutions of the Laplace equation~\cite{potential,green,meylan1996a,meylan2002,meylan2015bc,meylan2021,meylan2021good,pottheory}, reduces the problem to Fredholm integral equations of the second kind for surface density functions. This method has proven particularly effective for computing hydrodynamic forces, added mass coefficients, and damping coefficients needed to describe the floe's dynamic response~\cite{isaacson,books,fenton,squire,evans,john1,john2}.

Recent work has begun to address more complex scenarios. Khakimzyanov and Dutykh~\cite{2019dutykh} explored long wave interactions with partially immersed bodies using multiple approaches including fully nonlinear weakly dispersive models and Boussinesq-type equations. Evans~\cite{evans} developed theory for wave-power absorption by oscillating bodies, relevant to understanding energy dissipation mechanisms in ice-wave systems.

Despite these advances, the existing literature exhibits notable limitations. The vast majority of studies have focused exclusively on either large, thick, rigid floes where elastic deformations are negligible and only rigid body motions (heave, surge, pitch, roll) are considered, or on very thin, effectively massless floes where elastic flexural modes dominate but inertial effects are ignored. Furthermore, nearly all analytical treatments have been restricted to axisymmetric geometries with uniform thickness, primarily circular disks. While some recent work~\cite{meylan2002,x1,x2,x3} has attempted to address arbitrary geometries, these extensions remain limited and have not fully incorporated the effects of non-uniform thickness distributions combined with realistic mass properties and simultaneous rigid-elastic coupling. The present work addresses these gaps by developing a comprehensive model for ice floes of arbitrary shape and non-uniform thickness, treating both rigid body motions and elastic deformations on equal footing within a unified Green function framework.

In this paper we analyze the  dynamics of interaction between a floe of arbitrary shape (with low aspect ratio of the horizontal section) and non-uniform thickness, with elastic properties including flexural deformation, also considering the rigid motions of the floe   (heave, surge, roll, pitch).

The paper is organized as follows. In section \ref{sec2} we introduce the model consisting in a compact ice floe of arbitrary shape with low aspect ratio for the horizontal section. In subsection \ref{subsecbasic} we present the basic hypotheses of the model. The floe region denoted $R_E$ is considered as a linear elastic body, freely floating over a region of water denoted $R_F$. We do not limit in this model to floes of axial symmetric shapes, yet we do not consider floes with one size way larger than the others. The model can be used for any circular, or polygonal shape, or irregular shapes that can still be bounded by two concentric circles of close values for the radii. Moreover, we consider non-uniform thickness $d(r,\theta)$ floes where the thickness function is not necessary axial symmetric. The boundaries of regions $R_F, R_E$ are divided in sub-surfaces $S_B, S_F, S_{\sigma}, S_I$ representing  the rigid bottom, the free water surface, a vertical cylindrical surface considered an artificial boundary for the scattered solutions, and the interface between the floe and water, respectively. The water is considered ideal fluid in potential flow obeying the Laplace equation for the  potential. The degrees of freedom of the floe are identified and classified in four types of dominant rigid motions (heave, surge, roll, pitch labeled by $k=1,\dots,4$) and in the flexural normal modes of vibration of the floe, labeled by a multi-index $\alpha=(j,m)$. For each degree of freedom we introduce the corresponding flow potential, and their linear combination forms the solution for the water flow. In addition, we have the incident waves potential $\phi_{-1}$ whose characteristics (frequency, wave number, wave height) are the free parameters of the model.

In subsection \ref{subsecBC} we introduce all boundary conditions for this model. Namely, we use non-penetration free-slip boundary conditions at the water bottom, scattering conditions for the scattered potential $\phi_0$ at the interface, kinematic velocity matching conditions at the interface $S_I$, first-order absorbing boundary conditions at the artificial boundary $S_{\sigma}$, free water surface Bernoulli  conditions, and dynamical boundary conditions between the fluid pressure and the elastic pressure in the floe at the interface $S_I$.

In section \ref{secelamo} we discuss the elastic model for the floe using a  Kirchhoff-Love, non-uniform thickness, non-uniform flexural rigidity approach. We obtain the equation for the transverse deflection $w$ of the interface  $S_I$,  and for the pressure $P$. We use free edge boundary conditions, formulate the eigen-problem, and present the spectrum, and the orthogonal basis of the eigenfunctions, i.e. the normal modes $(j,m)$.

In section \ref{seccoupl} we describe   the coupling conditions for the  boundary conditions at the interface $S_I$ between the elastic modes and the fluid. We obtain for the flow partial potentials $\phi_{jm}$, the pressure $P$ at the interface and the transverse deflection  $w$ expansions in series of normal modes.
Using the previously described eigenproblem and both the kinematic and the dynamic boundary conditions, we eliminate the pressure and deflection from the system and obtain solutions  for the partial potentials. The time dependence is described by normal frequencies of vibration in vacuum,  corrected for the added mass effect of water. 

In section \ref{secbcrig} we introduce the boundary conditions associated to the floe rigid motions. We calculate the normal to the interface $S_I$ using differential geometry. We analyze each rigid motion degree of freedom and obtain exact analytic expressions, in the linear approximation of small oscillations for all these modes heave, surge, roll, and pitch.

The interactions between water and the dynamics of the rigid modes is developed in subsection \ref{subdybori}. We use the dynamical boundary conditions for such rigid motions and obtain expressions for generalized forces (i.e. forces and momenta) in the Bernoulli linear approximation. Using the conventional way for treating the motion of a floating body, by the added-mass coefficients and damping coefficients, we obtain the corresponding matrices and gather all these information into and algebraic  equation for the amplitudes of the rigid motions $\xi_k$.

The main result of the paper is presented in section \ref{secpotsol} where we obtain the general solution for the potential and the floe dynamics. We  re-write all the boundary conditions for all the sub-surfaces presented above in a compact tensor form, with coefficients classified in three tables. The Laplace equations can now be solved using  variable coefficient Robin type of boundary conditions for each degree of freedom. Using the single layer potential representation of solutions, and the Green function method we obtain the solutions for every partial potential in terms of integral representations, where the density functions obey  Fredholm types of integral equations of the second kind, for all degrees of freedom. In continuation, from the potential expressions we obtain solutions for the transverse deflection of the plate and the rigid motions.
We discuss time dependence for various incident waves of excitations and possible resonances. 

In Appendix A we present an example of analytic solution for the flexural eigenproblem with free edge boundary conditions under the assumption of an axis-symmetric top-down conical shape with a linear expression for the ice thickness, and we develop an asymptotic analysis to demonstrate calculation of the spectrum.

In Appendix B we present an example of calculation of the natural frequencies of elastic modes in vacuum with free edges boundary conditions for a disk of  uniform thickness.

\section{The model equations}
\label{sec2}

\subsection{Basic hypotheses}
\label{subsecbasic}

\begin{figure*}
	\centering
	\includegraphics[scale=.43]{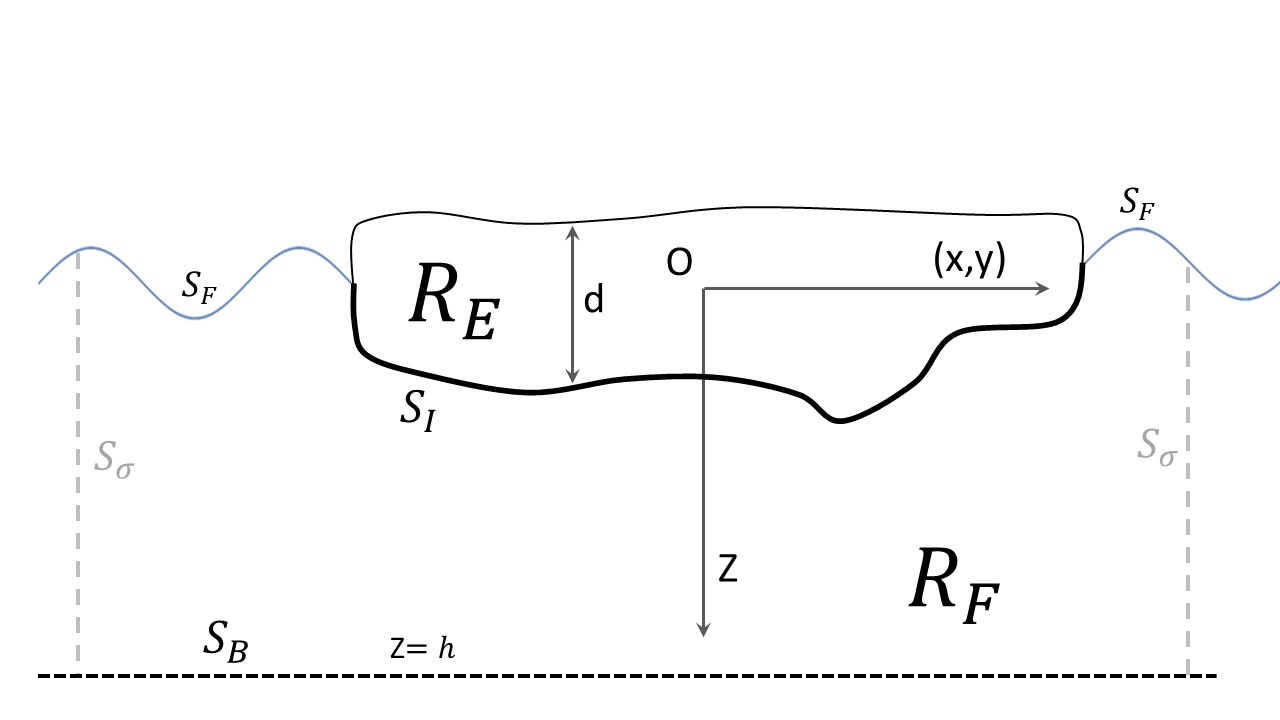}
	\caption{A quasi-planar ice floe region $R_E$ of non-uniform thickness $d$ floating over the water domain $R_F$ which is bounded by the free water surface $S_{F}$, interface with the floe $S_I$, horizontal water bed $S_B$ at $z=h$ and auxiliary cylindrical surface $S_{\sigma}$.}
	\label{Fig1}
\end{figure*}

We study a linear model of interaction between a deformable sea-ice floe and surface water waves.  The water  is considered incompressible (homogeneous e.g. salinity), inviscid  and the flow is irrotational with surface tension neglected. The surface waves are asymptotically linear long-crested (plane) monochromatic waves with small amplitude compared to the water depth. Tides, currents, convection motion,  winds,  and Coriolis force contributions are also neglected for the time and space scales considered. The ice floe  $R_E$ Fig. \ref{Fig1} is represented by a quasi-planar body (nonuniform lamina or planoid) laterally bounded by a nearly-circular contour, having homogeneous distribution of physical parameters inside, non-uniform thickness $d$, obeying linear elastic equations, and having its horizontal extension much larger than its average thickness. Phase changes at the ice floe surface are also neglected.  

In the present literature  \cite{isaacson,john1,john2,massonleblond,meylan1996a,meylan2002,meylan2015bc,meylan2021,meylan2021good,squire,x1,x2,x3}, the time dependence of all quantities is introduced in the Fourier representation, and occasionally a certain Fourier component is separated and used for the asymptotic region of the water waves which interact with the floe. In this case the velocity potential is thus expressed $\Phi(\vec{r},t)=\mathit{Re}[\phi(\vec{r}) \exp(-i \omega t)]$ with  $\vec{V}=\mathit{Re}[\nabla \Phi (\vec{r},t)]$ and $\triangle \phi=0$.  In this study, based on the linear approximation,  we still consider a separation of variables in the form space-time, but with general time dependence $\Phi(\vec{r},t)=\mathit{Re}[\phi(\vec{r}) \varphi(t)]$ with  $\vec{V}=\mathit{Re}[\nabla \Phi (\vec{r},t)]$ and $\triangle \phi=0$.

We consider a Cartesian system of coordinates with the $(x,y)-$plane parallel to the quiescent surface, and $z-$axis downwards, and when is the case we use the polar coordinates denoted $\vec{r}=(r, \theta, z)$ unless otherwise specified.
The water domain $R_F$ is bounded by a piece-wise smooth surface  $\Sigma=S_F \cup S_I \cup S_B \cup S_{\sigma}$ consisting in the free water surface $S_{F}$ where we can approximate $z=0$ for small amplitude waves,  the interface between water and the floe $S_{I}$, the horizontal bed $S_B$ placed at a depth $z=h>0$, and an auxiliary vertical cylindrical surface $S_{\sigma}$ of radius $\sigma$ added for computational truncation of a the semi-infinite region of the ocean, assumed to obey non-reflecting conditions Fig. \ref{Fig1}. Once a formal solution is obtained using the Green function method, the surface $S_{\sigma}$ can be expanded to infinity.  

Following \cite{isaacson,books,meylan2015bc,meylan2021good,meylan2021,x1,x2,x3} we consider the potential as a linear combination of independent partial potentials associated with the following scenarios: incident wave (or asymptotic potential) $\phi_{-1}$, scattered waves of the rigid surface $\phi_0$, and waves generated by the floe degrees of freedom of rigid motions $\phi_k$, $k=1,\dots,4$.  The most relevant independent degrees of freedom of rigid motion, when floe is excited with long-crested monochromatic waves along the $x$ direction, are: heave (translation along $z$) for $k=1$, surge (translation along $x$) for $k=2$, roll (rotation around $x$) $k=3$, and pitch (rotation around $y$) for $k=4$. To each such mode we associate an amplitude of motion $\xi_k$, $k=1,\dots,4$ and  we treat these degrees of freedom as small linear oscillations.

We add to these rigid motions degrees of freedom new independent degrees of freedom  for the elastic modes of deformation the floe (flexural modes) denoted $\phi_{\alpha}$, where $\alpha$ is a multi index determined by the spectrum and orthogonal basis of eigenfunctions of the elastic floe problem. 

Each such partial potential obeys the Laplace equation, and we have:
\begin{equation}
	\phi=\sum_{k=0}^{4}\xi_k \phi_k+\sum_{\alpha}\phi_{\alpha}, \ \ \triangle \phi_k=0, \ \ \triangle \phi_{\alpha}=0, \ \vec{r}\in R_F, 
	\label{lnbc11} 
\end{equation}
where we included in the partial potentials generated by rigid motions degrees of freedom the amplitude of motions $\xi_k$ \cite{massonleblond,isaacson,john1,john2,fenton,books}.

The incident potential is described in our model by linear long-crested waves, traveling along the $x$ direction \cite{isaacson}: 
\begin{equation}
	\phi_{-1}=-\frac{i g A}{2 \omega}\frac{\cosh [\kappa (z-h)]}{\cosh (\kappa h)}e^{i \kappa x},
	\label{lnbc16} 
\end{equation}
with $g$ gravitational constant, and where parameters $\kappa$ and $\lambda=2 \pi/\kappa$  are the wave number and the wavelength, respectively, associated to the Fourier component $\omega$ of the linear waves, and they are determined by the linear waves dispersion relation  $\omega=\sqrt{g \kappa \tanh(\kappa h)}$.

\subsection{Boundary conditions}
\label{subsecBC}

To Eqs. (\ref{lnbc11}) we associate two type of boundary conditions (BC). The first type, kinematic BC are given as follows. We use the usual non-flow condition at the bottom surface $S_B$:
\begin{equation}
	\frac{\partial \phi_k}{\partial z}\biggr|_{S_B}=0, \ k=0,\dots,4 \hbox{   and } \frac{\partial \phi_{\alpha}}{\partial z}\biggr|_{S_B}=0.
	\label{lnbc14}
\end{equation} 
The scattered waves $\phi_0$ are generated by total reflection of the incident potential $\phi_{-1}$ off the rigid surface $S_I$ considered fixed \cite{john1,john2,fenton,isaacson}: 
\begin{equation}
	\biggl( \frac{\partial \phi_0}{\partial \nu}+\frac{\partial \phi_{-1}}{\partial \nu}\biggr)_{S_I}=0, 
	\label{lnbc12} 
\end{equation}
where $\partial / \partial \nu$ is the normal derivative at surface $S_I$ of unit normal $\vec{\nu}$ in its equilibrium position, in absence of motions and flexural deformations.

The kinematic boundary conditions  at $S_I$ are obtained from matching the solid interface velocity with the fluid velocity, for each mode of motion and deformation, and for each partial potential, respectively. In the approximation of ideal fluid this boundary condition requests  that the normal component of the fluid velocity relative to the piston must be zero at the piston surface (no-penetration condition). The tangential component doesn't appear in this boundary condition because ideal fluids have no viscosity, so there is no constraint on tangential motion, and the fluid can freely slip along the surface. We have:
\begin{equation}
	\frac{\partial \phi_k}{\partial \nu_k}\biggr|_{S_I}=v_{\nu k}, \ k=1,\dots,4 \hbox{   and } \frac{\partial \phi_{\alpha}}{\partial \nu_{\alpha}}\biggr|_{S_I}=v_{\nu \alpha},
	\label{lnbc12b} 
\end{equation}
where $\vec{\nu}_k, \vec{\nu}_{\alpha}$ are the unit normal vectors to $S_I$ while in motion with mode $k$, or deformed under the mode $\alpha$. Here $v_{k}, v_{\alpha}$  are the velocities of the  surface $S_I$ corresponding to the degree of freedom $k,\alpha$, and $v_{\nu k}=\vec{v}_k\cdot \vec{\nu}_k, v_{\nu \alpha}=\vec{v}_{\alpha}\cdot \vec{\nu}_{\alpha}$. If we consider the ice surface $S_I$  rough and we consider water viscosity, there will be tangential shear forces, and the flow will include a non-potential component. In this case we need to use in addition to the normal boundary condition Eq. (\ref{lnbc12b}) a tangent,  no-slip boundary condition $(\vec{V}_k \times \vec{\nu}_k)_{S_I}=(\vec{v}_k \times \vec{\nu}_k)_{S_I}$. Near the rough surface, viscous effects create a boundary layer where velocity transitions from the $S_I$ velocity to the outer flow. The tangential velocity discontinuity at the $S_I$ surface generates vorticity, making the flow rotational near the ice interface. In this case we should use the Helmholtz-Hodge decomposition of the fluid velocity field and consider, besides the potential flow,  a vortical component for water velocity generated by the shear stress. In this study, we neglect this rotational component; however, such a component can always be added to our model.

Finally, at the artificial boundary $S_{\sigma}$ we impose a first-order absorbing boundary condition
\begin{equation}
\biggl( i \kappa \phi_{k}- \frac{\partial \phi_k}{\partial r}\biggr)_{S_{\sigma}}=0  \hbox{  for  } k=0,\dots,4 \hbox{  and } \biggl( i \kappa \phi_{\alpha}- \frac{\partial \phi_{\alpha}}{\partial r}\biggr)_{S_{\sigma}}=0,
	\label{lnbc15}
\end{equation}
which approximates the Sommerfeld behavior \cite{somm1,hist,x3,meylan2002} and it is generally valid if the radius $\sigma$  of the auxiliary cylindrical surface $S_{\sigma}$ is much greater than the typical wavelength of the incident water waves, and/or if the potentials are dominated by the radial oscillation (small angular variation). In principle this condition produces reflections (error depends on wavelength, angle of incidence, and $\sigma$), but in the case of not very wide and close to circular shape ice floes it is sufficient. Higher-order local conditions like Bayliss–Turkel \cite{bc1}, or Engquist–Majda \cite{bc2} reduce reflections by including angular derivatives and higher powers of $1/r$, yet again, in our case we do not need this degree of precision, especially since in practical cases $h,\sigma \rightarrow \infty$. Alternatively, we could use a nonlocal Dirichlet-to-Neumann \cite{dtn}
\begin{equation}
	\frac{\partial \phi_{k}}{\partial r}(\sigma,\theta)=\sum_{m=-\infty}^{\infty} \frac{\kappa H_{m}^{(1)'}(\kappa \sigma)}{H_{m}^{(1)}}\hat{\phi}_{k,m} e^{i m \theta}, \ \  k=0, \dots,4,
	\label{lnbc1}
\end{equation}
where $\hat{\phi}_{k,m}$ is the $m-$mode of the Fourier transform of the potential $\phi_k$ with respect to azimuth angle $\theta$. However, the implementation of the condition in Eq. (\ref{lnbc1}) requests truncation of the involved Fourier series, so for our geometry this condition is not more precise than Eq. (\ref{lnbc15}).

In the following, we implement the dynamical BC. For the free water surface we use the linearized Bernoulli equation coupled with the linearized kinematic free surface condition \cite{landa,john1,john2,books}:
\begin{equation}
\biggl( g\frac{\partial \phi_k}{\partial z}+\frac{\partial^2 \phi_k}{\partial t^2}\biggr)_{S_F}=0 \hbox{   where } k=0,\dots,4 \hbox{  and } \biggl( g\frac{\partial \phi_{\alpha}}{\partial z}+\frac{\partial^2 \phi_{\alpha}}{\partial t^2}\biggr)_{S_F}=0 . 
	\label{lnbc13} 
\end{equation}
The most complex dynamical BC occurs at the interface $S_I$ because of the coupling between the motions and deformation modes of the elastic body and the hydrodynamic pressure underneath. Following \cite{landa,evans,x1,x2,x3,meylan2002,meylan2021,meylan2021good} we have for each flexural mode $\alpha=(j,m)$, again from the linearized Bernoulli equation
\begin{equation}\label{e100}
	P_k\biggr|_{S_I}= -\rho \frac{\partial \phi_k}{\partial t}\biggr|_{S_I}, \ k=1,\dots,4 \ \ \hbox{and }  \ P_{\alpha}\biggr|_{S_I}=\biggl( -\rho \frac{\partial \phi_{\alpha}}{\partial t}-\rho g w_{\alpha} \biggr)_{S_I}, 
\end{equation}
where $\rho$ is the density of water, and $w_{\alpha}$ is the local vertical floe displacement associated to the flexural mode ${\alpha}$, to be  defined rigorously in the next section.

\section{The elastic model for the floe}
\label{secelamo}

In order to evaluate the pressure response of the elastic floe we use the F\"{o}ppl-von K\'{a}rm\'{a}n  elastic plate model for non-uniform thickness \cite{flop,lei,love,sode}, yet with uniform distribution of the other mechanical and elastic parameters. We describe the floe interface with water $S_I$ with a function $r=R(\theta,z, t)$, and the thickness of the floe a function $d(r,\theta)$. In our model, since we consider the floe as a quasi-planar elastic body, we need to elaborate on  the transverse (vertical) displacement $w(r,\theta,t)$ of the interface $S_I$. From the point of view of the interaction with water, we take into account the arbitrary geometry of the interface $S_I$ through $r=R(\theta, z,t)$, but in the study of the elastic deformation of the floe we consider only its transverse displacement as the only independent variable. 

For the elastic bending part of the model, the most general governing equation for the transverse displacement $w(r,\theta,t)$, has the form \cite{lei,la,lu,sode}
\begin{equation}\label{e101}
	\nabla \cdot [D \nabla (\triangle w)]-\nabla D \cdot \nabla (\triangle w)+\rho_{e} d \frac{\partial^2 w}{\partial t^2}=P, \hbox{  on } S_I.
\end{equation}
where $D$ is the non-uniform flexural rigidity 
\begin{equation}\label{ddd}
	D(r,\theta)=\frac{E d^3(r, \theta)}{12 (1-\nu_{e}^{2})},
\end{equation}
and $P(r, \theta, t)$ is the pressure of the water on the interface $S_I$. Here $\rho_e$, $E$, $\nu_{e}$ are the floe density, Young's modulus, and Poisson's ration for the floe, respectively, and have uniform values through its volume $R_E$. The first term in the left term represents the biharmonic operator with variable coefficients from the Kirchhoff-Love model, and the second term represents curvature coupling terms \cite{lei,flop,sode}.  

For floes of lateral size  in the range from tens to hundreds of meters, thickness in the range $1$m $ < d < 3$m and typical ocean long-crested waves with wavelength in the range of tens of meters, we have a very small gradual thickness variation $\epsilon =(d_{max}-d_{min})/d_{mean} < 10^{-2}$, meaning that the thickness changes in average around $1\%$ per $5-10$m radially.  Ocean waves with wavelengths comparable to or larger than the floe dimension (typical for long-crested waves) also produce smooth, large-scale deflection patterns. The bending wavelength is much larger than the thickness variation scale. In addition we consider a small deflection regime in the range $w \sim \mathcal{O}(0.1-1)$m giving $w/d \sim 0.1-1$, which is a small deflection regime. Consequently,  since the geometric nonlinearity remains relatively weak, you can safely neglect curvature coupling terms $\nabla D \cdot \nabla (\triangle w)$ in a first-order linear analysis. In the following, we use the generalized Kirchhoff-Love model
for non-uniform thickness \cite{h,love,flop,sode}
\begin{equation}\label{e101b}
	\nabla \cdot [D \nabla (\triangle w)]+\rho_{e} d \frac{\partial^2 w}{\partial t^2}=P, \hbox{  on } S_I.
\end{equation}
In the generalized Kirchhoff-Love model the boundary conditions on the free edge of the elastic floe  \cite{h,love,lei,flop,sode,senjan,meylan2002,x1,x2,x3} are given by annihilating the bending moment in the normal direction, at the edge $r=R(\theta, 0, t)$:
\begin{equation}\label{e102}
	M_{\nu}=-D\biggl[ \frac{\partial^2 w}{\partial \nu^2}+\nu_e \triangle w \biggr]=0,
\end{equation}
and also by annihilation of the effective shear (Kirchhoff effective shear) at the edge:
\begin{equation}\label{e103}
	V_{\nu}=-\frac{\partial }{\partial \nu} (D \triangle w)-\frac{\partial }{\partial s}\biggl[ D (1-\nu_e ) \frac{\partial^2 w}{\partial \nu \partial s}\biggr] =P_{\nu} (\theta, t),	
\end{equation}
where $\nu$ and $s$ represent the normal and tangent directions to $S_I$, respectively, and $P_{\nu}$ is the applied normal pressure distribution. 

In the present model, we neglect in Eq. (\ref{e103}) the contribution of the normal pressure $P_{\nu}$ applied by water on the edges of the floe, that is, for a narrow region of $S_I$ given by $r=R(\theta, z,t)$ with $0<z<w$. The reason is straightforward: the boundary conditions Eqs. (\ref{e102}, \ref{e103}) are used to complete a certain eigen-problem in order to build an  orthogonal basis for the modal expansion of the transverse displacement $w(r, \theta, t)$. So the normal pressure on the floe edge has contribution only for the normal modes of vibration of the flow, that is for the degrees of freedom $\phi_j$ coupled with the floe vibrations. Nevertheless, for relatively large and thick floes the contribution of the normal modes of vibration at the edge is pretty negligible compared to the contribution of the solid movements of the flow (surge, heave, pitch) on the edge.  Consequently we can use the boundary condition of annihilation of the effective shear Eq. (\ref{e103}) as a homogeneous boundary condition at the edge:
\begin{equation}\label{e103x}
	V_{\nu}=-\frac{\partial }{\partial \nu} (D \triangle w)-\frac{\partial }{\partial s}\biggl[ D (1-\nu_e ) \frac{\partial^2 w}{\partial \nu \partial s}\biggr] =0.	
\end{equation}

Eqs. (\ref{e101b}, \ref{e102}, \ref{e103x}) form highly coupled system where the non-uniform thickness $h(r,\theta)$ creates a strong coupling between radial and azimuthal modes through the gradient terms of $D$. In the case of uniform thickness, Eqs. (\ref{e101}, \ref{e101b}]) reduce to the Kirchhoff-Love displacement equation for uniform thick plates \cite{h,love,senjanx1,x2,x3,meylan2002}.

Eq. (\ref{e101b}) coupled with appropriate homogeneous boundary conditions Eqs. (\ref{e102}, \ref{e103x}), and considering the variable coefficients $d,D$ to be smooth enough functions,  provide  a 2-dim self-adjoint elliptic eigenvalue problem \cite{lei,sode}, that shares key properties with any Sturm-Liouville Problem, like real eigenvalues $\lambda_{jm}$, orthogonal eigenfunctions $w_{jm}$, completeness of eigenfunctions, and variational characterization (Rayleigh quotient) \cite{flop,h,fol,free}, in the form
\begin{equation}\label{e101c}
	\mathcal{L}[w_{jm}]=\nabla \cdot [D \nabla (\triangle w_{jm})]=\lambda_{jm} \rho_{e} d \ w_{jm},
\end{equation}
where $\mathcal{L}$ is a self-adjoint, biharmonic, variable-coefficient fourth-order operator. The eigenfunctions $w_{jm}$ are orthogonal with respect to the weighted inner product 
$$
\int_0^{2 \pi}\int_0^{R(\theta ,0,0)}w_{jm}(r, \theta)w_{j'm'}(r, \theta) \rho_e d(r, \theta) \ dr \ d\theta=0,  
$$
for $(j,m)\neq(j',m')$. The difficulty involved by this general treatment of the floe is that, compare to previous literature case studies  \cite{isaacson,john1,john2,massonleblond,x1,x2, x3,meylan1996a,meylan2002,meylan2015bc,meylan2021,meylan2021good,squire}, one  cannot separate variables into $R(r)\Theta(\theta)$ anymore. This means that there is no more a clean decomposition into radial and angular modes, and the $(j,m)$ numbers lose their simple interpretation of  radial and circumferential mode numbers. The separation of variables and restoration of  modes meaning can still work in the non-uniform thickness case only if the floe has circular symmetry, that is if the thickness function is described by $d=d(r)$. 

An example of asymptotic calculation of this eigen-problem, for axis-symmetric floe, is presented in Appendix A.  In the general thickness case the self-adjointness and orthogonality properties persist, but the practical utility of modal decomposition into radial and circumferential modes is lost.  Most importantly, we still have a complete orthogonal basis, even if the  modes are fully 2-dim.

\section{Coupling the  boundary conditions for the elastic deformation modes on the interface $S_I$}
\label{seccoupl}

In this section we discuss the partial potential induced by the elastic deformation of the floe. We follow the same mathematical procedure and steps as in \cite{x1,x2, x3,meylan1996a,meylan2002,meylan2015bc,meylan2021good,meylan2021}, but for our general case. For all the functions defined on the surface $S_I$ we use the 2-dimensional $(r, \theta)$ modal expansion  resulting from the eigen-problem Eqs. (\ref{e102}, \ref{e103x}, \ref{e101c}) for the elastic deformation of the floe, because these modes form a complete orthogonal basis as described in the previous section. In the following, we solve the system formed by all boundary conditions at the  interface $S_I$, namely Eqs. (\ref{lnbc12b}, \ref{e100}, \ref{e101b}). We assume the following modal expansions for the partial potentials and pressure  on $S_I$, and for the transverse displacement
$$
\phi_{flexural}|_{S_I} = \sum_{\alpha}\phi_{\alpha} = \sum_{j,m} \psi_{jm}(r, \theta) a_{jm}(t),
$$
$$
P|_{S_I} = \sum_{j,m} \Pi_{jm}(r, \theta) p_{jm}(t),
$$
\begin{equation}\label{m1}
	w = \sum_{j,m} w_{jm}(r, \theta) b_{jm}(t),
\end{equation}
where $w_{jm}$ are the orthogonal basis of eigenfunctions from Eq. (\ref{e101c}). We implement the modal series from Eqs. (\ref{m1}) in the boundary conditions Eqs. (\ref{lnbc12b}, \ref{e100}, \ref{e101b}) and obtain, for each $j,m$ respectively:
\begin{equation}\label{m2}
	\frac{\partial \psi_{jm}}{\partial \nu_{jm}}a_{jm}=w_{jm}   b'_{jm},
\end{equation}
\begin{equation}\label{m3}
	\rho  (\psi_{jm} a'_{jm} +g w_{jm} b_{jm})=- \Pi_{jm} p_{jm},
\end{equation}
\begin{equation}\label{m4}
	\mathcal{L} [  w_{jm}] b_{jm}+\rho_e d  w_{jm} b''_{jm}= \Pi_{jm} p_{jm}.
\end{equation}
We combine these equations to eliminate $P,w$ and obtain one equation in terms of $\phi$ only, which will also provide the characteristic equation for the time dependence. We plug Eq. (\ref{m2}) in the time differentiated Eq. (\ref{m3}) and obtain
\begin{equation}\label{m5}
	\rho  \biggl( \psi_{jm} a^{''}_{jm} +g \frac{\partial \psi_{jm}}{\partial \nu_{jm}}a_{jm}\biggr) =- \Pi_{jm} p^{'}_{jm},
\end{equation}
Next, Eq. (\ref{m5}) and time derivative of Eq. (\ref{m4}) we can eliminate $\Pi_{jm}$
\begin{equation}\label{m6}
	-\rho  \biggl( \psi_{jm} a^{''}_{jm} +g \frac{\partial \psi_{jm}}{\partial \nu_{jm}}a_{jm}\biggr) =\mathcal{L}\biggl[  \frac{\partial \psi_{jm}}{\partial \nu_{jm}} \biggr] a_{jm} +\rho_e d \frac{\partial^2}{\partial t^2}\frac{\partial \psi_{jm}}{\partial \nu_{jm}}a_{jm}.
\end{equation}
This equation can be factored under the hypothesis $a^{''}_{jm}=-\tau^{2}_{jm} a_{jm}$ for constant $\tau_{jm}$, resulting in solutions for the time dependent part 
\begin{equation}\label{m7}
	a_{jm}(t)=a^{0}_{jm} e^{\pm i \tau_{jm} t} \hbox{   and  } b_{jm}(t)=b^{0}_{jm} e^{\pm i \tau_{jm} t}.  
\end{equation}
Using again Eq. (\ref{m2}) and the eigen-problem Eq. (\ref{e101c}) we can re-write Eq. (\ref{m6}) in the form
\begin{equation}\label{m8}
	-\rho  \biggl(-\tau^{2}_{jm} \psi_{jm} a^{0}_{jm} +g \frac{\partial \psi_{jm}}{\partial \nu_{jm}}a^{0}_{jm}\biggr) =\pm i  \tau_{jm} \lambda_{jm} \rho_e d  w_{jm} b^{0}_{jm} -\rho_e d  \tau^{2}_{jm} \frac{\partial \psi_{jm}}{\partial \nu_{jm}}a^{0}_{jm}.
\end{equation}
or simply
\begin{equation}\label{m9}
	\psi_{jm}   =\pm i \frac{b^{0}_{jm}}{a^{0}_{jm}}  \biggl( \frac{g}{ \tau_{jm}}   +  d\frac{\lambda_{jm}}{\tau_{jm}}  \frac{\rho_e}{\rho}  +  d \tau_{jm}  \frac{\rho_e}{\rho}  \biggr) w_{jm}.
\end{equation}
This expression confirms that the partial potential functions on the interface $S_I$ are proportional to the eigenfunctions of the well posed flexural problem. For floes of uniform thickness ($d=$constant) the partial potentials are proportional to the flexural eigenfunctions. From Eqs. (\ref{m1}, \ref{m9}) we have the value of the contribution to the potential of flexural deformations in the interface $S_I$
\begin{equation}\label{m10}
	\phi_{flexural}|_{S_I} =\pm i \sum_{j,m}  b^{0}_{jm}  \biggl( \frac{g}{ \tau_{jm}}   +  d\frac{\lambda_{jm}}{\tau_{jm}}  \frac{\rho_e}{\rho}  +  d \tau_{jm}  \frac{\rho_e}{\rho}  \biggr) w_{jm} e^{\pm i \tau_{jm} t}.
\end{equation}
Concerning the frequencies $\tau_{jm}$  we  can make the assumption that they are given by the modified  natural frequency modes which take into account the added mass effect from the fluid. This is actually a fundamental aspect of the modal decomposition approach for fluid-structure interaction problems. Each partial mode $\psi_{jm}$  oscillates harmonically at the corresponding modified natural frequency. The corrections for the added mass effect of water for the elastic floe are given by \cite{ew1,ew2,ew3} 
\begin{equation}\label{etau}
\tau^{2}_{jm}=\frac{\omega^{2}_{jm}}{1+\frac{m_{add, jm}}{m_{floe}}}
\end{equation}
where $\omega_{jm}$ are the natural frequencies of mode $(j,m)$ in vacuum, see Appendix B. The coefficients $m_{add, jm}$ can be calculated using the usual formula \cite{ew1,ew2,ew3,ew4,ew5,ew6}
$$
m_{add, jm}=\rho \iint_{S_I}\psi_{jm}(r, \theta) w_{jm} (r, \theta) \ dA
$$
where we substitute for $\psi_{jm}(r, \theta)$ the expression from Eq. (\ref{m9}), resulting in an implicit algebraic equation for $\tau_{jm}$. Next, the partial potentials on $S_I$ for each normal mode are calculated using Eq. (\ref{m9}), and then the pressure on the
interface $S_I$ can be obtained from any of the last two equations in Eq. (\ref{m3}).

\section{Boundary conditions associated to rigid motions}
\label{secbcrig}

For the modes corresponding to the rigid motions of the floe $\xi_k \phi_k$, $k=1,\dots,4$ we need to evaluate  the kinematic boundary conditions at the interface $S_I$ assumed rigid and in motion Eq. (\ref{lnbc12b}), and the dynamic boundary conditions Eq. (\ref{e100}). We parametrize the interface $S_I$ in cylindrical coordinates
$\vec{r}_{S_I}=(r, \theta, d(r, \theta))$ where $d$ is the non-uniform thickness function. The outer unit normal $\vec{\nu}$ to this surface and its projections on the $x-$ and $z-$axes are given by differential geometry
$$
\vec{\nu}=\frac{-d_r \vec{e}_x -\frac{d_{\theta}}{r} \vec{e}_{\theta}+ \vec{e}_z}{\sqrt{d_r^2 +\frac{d_{\theta}^2}{r^2}+1}}
$$
\begin{equation}\label{m11}
	\nu_x=\frac{-d_r \cos \theta}{\sqrt{d_r^2 +\frac{d_{\theta}^2}{r^2}+1}}, \ \ \ \nu_z=\frac{1}{\sqrt{d_r^2 +\frac{d_{\theta}^2}{r}+1}},
\end{equation}
where $d_r=\partial d/\partial r, d_{\theta}=\partial d/\partial \theta$ and $\{ \vec{e}_r, \vec{e}_{\theta} \}$ are the radial and azimuthal unit vectors.

In the following we use the approach introduced in \cite{isaacson}, where the normal velocities of $S_I$ from Eq. (\ref{lnbc12b}) are obtained from geometric transformations, corresponding for each rigid motion labeled by $k=1, \dots,4$, as we introduced them in the beginning of subsection \ref{subsecBC}.

\subsection{Heave mode kinematic boundary conditions, $k=1$}
\label{subsecheave}

We assume an oscillating horizontal motion of the floe in the $x-$direction, with amplitude $\xi_1$ (which for the moment is a free parameter) and frequency $\omega$.  For the mode $k=1$ the boundary condition Eq. (\ref{lnbc12b}) becomes
\begin{equation}
	\frac{\partial \phi_1}{\partial \nu_x}\biggr|_{S_I}=\nu_x \vec{e}_x\cdot \nabla \phi_1=i \omega  e^{i \omega t}
	\label{m12} 
\end{equation}
or explicitly
\begin{equation}
	\frac{-d_r \cos \theta}{\sqrt{d_r^2 +\frac{d_{\theta}^2}{r^2}+1}} \biggl( \cos \theta \frac{\partial \phi_1}{\partial r}-\frac{\sin \theta }{r}\frac{\partial \phi_1}{\partial r} \biggr) \biggr|_{S_I} =i \omega   e^{i \omega t}
	\label{m13} 
\end{equation}

\subsection{Surge mode kinematic boundary conditions, $k=2$}
\label{subsecsurge}

For the surge motion we also assume an oscillating vertical  motion of the floe in the $z-$direction, with amplitude $\xi_2$ and frequency $\omega$.  For the mode $k=2$ the boundary condition Eq. (\ref{lnbc12b}) becomes
\begin{equation}
	\frac{\partial \phi_2}{\partial \nu_2}\biggr|_{S_I}=\nu_z \vec{e}_z \cdot \nabla \phi_2=i \omega  e^{i \omega t}
	\label{m14} 
\end{equation}
or explicitly
\begin{equation}
	\frac{1}{\sqrt{d_r^2 +\frac{d_{\theta}^2}{r}+1}}  \frac{\partial \phi_2}{\partial z} \biggr|_{S_I} =i \omega   e^{i \omega t}
	\label{m15} 
\end{equation}

\subsection{Roll mode kinematic boundary conditions, $k=3$}
\label{subsecroll}

For the roll motion we  assume small oscillating rotational motion of the floe around the  $x-$direction, with angular amplitude $\xi_3$ and frequency $\omega$, where we  assume small linear oscillations approximations. The small small linear oscillation approximation involves  $\sin \xi_3 \simeq \xi_3, \cos \xi_3 \simeq 1$. The difference between this mode and translational modes is that each point and the unit normal $\vec{\nu}$ on the interface $S_I$ changes its position and orientation, respectively,  according to the rotation transformation. The parameterization of the interface $S_I$ changes accordingly
\begin{equation}\label{m18}
	\vec{r}^{\ '}=(r', \theta', z')= \biggl(\sqrt{r^2 \cos^2 \theta +( r \sin \theta +z \xi_3)^2}, \arctan \frac{r \sin \theta+z \xi_3}{r \cos \theta}, -r \xi_3 \sin \theta  +z \biggr).
\end{equation}
The unit normal at the new points on the interface is also rotated into $\vec{\nu}^{\ '}_{3}(\vec{r}^{\ '})$ given by
$$
(\nu_x, \nu_y, \nu_z) \rightarrow \vec{\nu}^{\ '}_{3}=(\nu_x,  \nu_y+ \xi_3 \nu_z, -\xi_3 n_y+\nu_z)
$$
being evaluated at rotated points $\vec{r}^{\ '}$. The direction of rotation of the points $\vec{r}^{\ '}$, which is the tangent direction to rotation, is given in Cartesian coordinates by $\vec{T}_{\xi_3}=(0, \xi_3, 1)$. The normal velocity of the points of $S_I$ in this rotation is given by
$$
v_{\nu 3}=i \omega \xi_3  (\vec{T}_{\xi_3}\cdot \vec{\nu}^{\ '}_{3}) \sqrt{r^2 \sin^2 \theta +d^2(r, \theta)}.
$$
The result is that for mode $k=3$ the boundary condition Eq. (\ref{lnbc12b}) becomes
\begin{equation}
	\frac{\partial \phi_3}{\partial \nu_3}\biggr|_{S_I}=\vec{\nu}^{\ '}_{3} \cdot \nabla \phi_3=i \omega   (\vec{T}_{\xi_3}\cdot \vec{\nu}^{\ '}_{3}) \sqrt{r^2 \sin^2 \theta +d^2(r, \theta)}  e^{i \omega t}.
	\label{m19} 
\end{equation}

\subsection{Pitch mode kinematic boundary conditions, $k=4$}
\label{subsecpitch}

For the pitch motion we also assume an oscillating rotational motion of the floe around the  $y-$direction, with angular amplitude $\xi_4$ and frequency $\omega$. We assume small linear oscillations so we can approximate $\sin \xi_4 \simeq \xi_4, \cos \xi_4 \simeq 1$. For this mode we also consider that the point and the unit normal $\vec{\nu}$ on the interface $S_I$ changes their position and orientation according to the rotation transformation. The parametrization of the interface $S_I$ in the fixed system of reference in quiescent water changes into a new equation  $\vec{r}^{\ '}=\hat{\mathcal{R}}(\xi_4)\vec{r}$ when the floe rotates, according to the rule for small angle approximation
\begin{equation}\label{m16}
	\vec{r}^{\ '}=(r', \theta', z')= \biggl(\sqrt{(r \cos \theta +z \xi_4)^2+r^2 \sin^2 \theta}, \arctan \frac{r \sin \theta}{r \cos \theta+z \xi_4}, -r \xi_3 \cos \theta  +z \biggr).
\end{equation}
The unit normal at the new points on the interface is also rotated into $\vec{\nu}^{\ '}_{4}(\vec{r}^{\ '})$ given by
$$
(\nu_x, \nu_y, \nu_z) \rightarrow \vec{\nu}^{\ '}_{4}=(\nu_x +\nu_z \xi_4, \nu_y, -\xi_4 \nu_x+\nu_z)
$$
being evaluated at rotated points $\vec{r}^{\ '}$. The direction of rotation of the points $\vec{r}^{\ '}$, which is the tangent direction to rotation, is given in Cartesian coordinates by $\vec{T}_{\xi_4}=(-\xi_4, 0, 1)$. The normal velocity of the points of $S_I$ in this rotation is given by
$$
v_{\nu 4}=i \omega s \xi_4  (\vec{T}_{\xi_4}\cdot \vec{\nu}^{\ '}_{4}) 
$$
where $s$ is the radius of rotation of any $S_I$ point $\vec{r}$ into $\vec{r}^{\ '}$ for the pitch mode is $s=\sqrt{r^2+d^2(r,\theta)}$.

The result is that for mode $k=4$ the boundary condition Eq. (\ref{lnbc12b}) becomes
\begin{equation}
	\frac{\partial \phi_4}{\partial \nu_4}\biggr|_{S_I}=\vec{\nu}^{\ '}_{4} \cdot \nabla \phi_4=i \omega s   (\vec{T}_{\xi_4}\cdot \vec{\nu}^{\ '}_{4})  e^{i \omega t}.
	\label{m17} 
\end{equation}

\subsection{Dynamical boundary conditions for rigid motions}
\label{subdybori}

We introduce the concept of generalized force acting on the floe  \cite{isaacson} defined either by a force acting on a specific direction, or a momentum with respect to a specific axis and the center of mass. To calculate the wave generalized forces upon the flow for the rigid motions degrees of freedom, we use Eq. (\ref{e100}) for $k=1,\dots,4$ and integrate the pressure over the interface. Similar to the previous section we consider all rigid degrees of freedom described by oscillations of frequency $\omega$. We have 
\begin{equation}\label{m20}
	F_{jk}=i \omega \rho e^{i \omega t} \int_{S_I} \phi_k \nu_j \ dS ,
\end{equation}
where $k=1,\dots,4$ describes the type of degree of freedom of rigid motions. The label $j=1,\dots,4$ describes the type of generalized force as follows: for $j=1$ we have the vertical $z-$component of the force; for $j=2$ we have the horizontal $x-$component of the force; for $j=3$ we have the momentum around the $x-$axis, and  for $j=4$ we have the momentum about the $y-$axis. According to \cite{books, fenton,isaacson} the generalized force components $F_{jk}$ are associated with the corresponding partial potentials and are conveniently described in the conventional way by the added-mass coefficients $\mathcal{A}_{jk}$, and damping coefficients $\mathcal{B}_{jk}$, both real coefficients, such that  
\begin{equation}\label{m21}
	F_{jk}=\omega^2 \mathcal{A}_{jk}-i \omega \mathcal{B}_{jk} \hbox{   or   } \mathcal{A}_{jk}=\frac{\mathbf{Re}(F_{jk})}{\omega^2}, \ \ \mathcal{B}_{jk}=-\frac{\mathbf{Im}(F_{jk})}{\omega}.  
\end{equation}
The equations of motion for a freely floating body of quasi-planar shape, laterally bounded by a nearly-circular contour, having homogeneous distribution of physical parameters inside, and non-uniform thickness, oscillating with four degrees of freedom may be written as \cite{yeung,isaacson,fenton,squire,evans,john1,john2}
\begin{equation}\label{m22}
	\sum_{j=1}^4 [-\omega^2 (\mathcal{M}_{kj}+\mathcal{A}_{kj})+i \omega \mathcal{B}_{kj}+\mathcal{C}_{kj}] \xi_j =F_{k}^{(e)},
\end{equation}
where $\mathcal{M}_{kj}$ are the mass matrix components, $\mathcal{C}_{kj}$ are the hydrostatic stiffness coefficients, and $F_{k}^{(e)}$ are the components of the exciting generalized force associated with the incident and scattered partial potentials $\phi_{-1}+\phi_{0}$, basically representing the wave generalized force for a fixed floe \cite{isaacson}.  These matrices are well-known in literature \cite{yeung,john1,john2,evans,squire,fenton,books}, and are defined below. The mass matrix has two terms:
$$
\mathcal{M}_{jk}=\mathcal{M}_{jk}^{rig}+\mathcal{M}_{jk}^{add}
$$
representing the rigid body mass and the added mass contributions. The rigid body mass matrix has the form
\begin{equation}\label{m23}
	(\mathcal{M}_{jk}^{rig})=
	\begin{bmatrix}
		m & 0 & 0 & 0 \\
		0 & m & 0 & m z_G \\
		0 & 0& I_{44} & -I_{45} \\
		0& m z_G &-I_{45}&I_{55}
	\end{bmatrix}
\end{equation}
where $m$ is the floe mass, $z_G$ is the vertical distance from origin (water surface) to the center of gravity of the floe, $I_{44}$ is  roll moment of inertia about $x-$axis through the center of gravity, $I_{55}$ is the pitch moment of inertia about $y-$axis through the center of gravity, and $I_{45}$ is the product of inertia \cite{books}.

The added mass matrix has the general rigorous form
\begin{equation}\label{m24}
	(\mathcal{M}_{jk}^{add})=\rho \iint_{S_I} \phi_j \nu_k \ dS.
\end{equation}
These coefficients have  complicated expressions, and in general they can be calculated only numerically. If floe has a shape closed to an axisymmetric shape, we can use for the added mass terms the approximation of an axisymmetric body. In this case the off-diagonal terms are neglected.

The hydrostatic stiffness matrix has the form \cite{books, isaacson}
\begin{equation}\label{m25}
	\mathcal{C}_{kj}=
	\begin{bmatrix}
		0 & 0 & 0 & 0 \\
		0 & \rho g A_w & \rho g A_w x_F & -\rho g A_w y_F \\
		0 & \rho g A_w x_F& \rho g A_w x_F^2+\rho g V z_B GM_T & -\rho g A_w x_F y_F \\
		0& -\rho g A_w y_F &-\rho g A_w x_F y_F&\rho g A_w y_F^2+\rho g V z_B GM_L
	\end{bmatrix}
\end{equation}
where $A_w$ is the  waterplane area at equilibrium, $V$
is the  displaced volume, $(x_F, y_F)$ are the coordinates of waterplane area centroid, and $z_B$ is the vertical distance from origin to center of buoyancy. The other two parameters are called the metacentric heights \cite{books} and are given by 
$$
GM_T=\frac{1}{V}\iint_{S_I} A_w y^2 \ dS-z_B+z_G
$$
for the transverse metacentric height, and 
$$
GM_L=\frac{1}{V}\iint_{S_I} A_w x^2 \ dS-z_B+z_G
$$
for the longitudinal metacentric height, where $z_G$ is the vertical distance from origin to center of gravity.

Finally for this section, the non-homogeneous term $F_{k}^{(e)}$ in the linear system Eq. (\ref{m22}) which represents the exciting generalized forces associated to the incident and scattered partial potentials can be obtained from the Haskind relations \cite{isaacson, books}. Instead of calculating this term using the general form Eq. (\ref{m20}) for $\phi_{-1}+\phi_{0}$ we can approximate this term with the particular expression for axisymmetric bodies \cite{isaacson} 
$$
F_{k}^{(e)}=\sqrt{
	K_k \frac{\rho g H^2 \mathcal{B}_{kk} \omega}{\kappa^2} \biggl[ 1+\frac{2 \kappa h}{\sinh ( 2 \kappa h)}\biggr]}
$$
in which $K_1=K_3=K_4=1$, $K_2=1/2$,height,  $H$ is the crest-to-crest exciting waves height, and $\kappa$ is the wave number associated to long-crested linear waves of frequency $\omega$ at depth $h$.

With all coefficients calculated as above, and using the linear approximation for the expressions of the rotated tangent and normal unit vectors 
$\vec{T}_{\xi_3}, \vec{T}_{\xi_4}, \vec{\nu}_3^{\ '}, \vec{\nu}_{4}^{'}$, 
we can solve the linear system Eq. (\ref{m22}) and obtain the values for the amplitudes of the rigid motions $\xi_k$ for some given, imposed exciting waves frequency $\omega$.

\section{The general solution for the potential and the floe dynamics}
\label{secpotsol}

In order to calculate the total potential in this problem we use the formal Green integral formula for the solution of Laplace's equation in the region $R_F$ with a Robin boundary condition on the closed surface $\Sigma$  \cite{fol,free,green,potential,h,pottheory,dtn,bc2}. The total potential Eq. (\ref{lnbc11}) has the form
$$	
\phi=\sum_{k=0}^{4}\xi_k \phi_k+\sum_{\alpha}\phi_{\alpha},
$$
where the parameters $\xi_k$ are calculated in section \ref{subdybori} as solutions of the system Eq. (\ref{m22}). The boundary condition is defined over the closed surface $\Sigma=S_F \cup S_I \cup S_B \cup S_{\sigma}$ in the form
\begin{equation}\label{m26}
\biggl( U_{\alpha \beta} \phi_{\alpha} + Z_{\alpha \beta} \frac{\partial \phi_{\alpha}}{\partial \nu_{\alpha \beta}} \biggr)_{S_{\beta}}=\Lambda_{\alpha\beta}
\end{equation}
where the indices run as follows: $\alpha=k$ or $(j,m)$ and $\beta=1,\dots,4$ labels the 4 surfaces forming $\Sigma$, or in other words $S_{\beta}\in \{  S_F, S_I,  S_B,  S_{\sigma} \}$. The coefficient functions $U_{\alpha \beta}, Z_{\alpha \beta}, \Lambda_{\alpha \beta}$ have their expressions determined in the previous sections, respectively. From the boundary conditions expressed in the general form Eqs. (\ref{lnbc14}-\ref{lnbc15}, \ref{lnbc13}, \ref{e100}), and the specific evaluated relationships in Eqs. (\ref{m10}, \ref{m12}-\ref{m15}, \ref{m19}, \ref{m17})
we obtain the following global structure for the functional coefficients $U, Z, \Lambda$ in the boundary condition Eq. (\ref{m26}):
\begin{table}[h!]
\centering
\begin{tabular}{| c | c | c | c | }
\hline
\multicolumn{4}{| c |}{$U_{\alpha \beta}$} \\
\hline
$\beta_{\downarrow}$  & $\alpha=0$ & $\alpha=k$ & $\alpha=(j,m)$ \\ \hline \hline
$S_F$ & $-\omega^2$ & $-\omega^2$ & $-\tau^{2}_{jm}$ \\ \hline
$S_I$ & $0$ & $0$ & $1$ \\  \hline
$S_B$ & $0$ & $0$ & $0$ \\ \hline
$S_{\sigma}$ & $-i \kappa$ & $-i \kappa$ & $-i \kappa$ \\
\hline
\end{tabular}
\caption{Table of values for the  functional coefficient $U_{\alpha \beta}$. Here $k=1,\dots,4$,   $\omega=\sqrt{g \kappa \tanh(\kappa h)}$, and the normal frequency coefficients $\tau_{jm}$ are given in Eq. (\ref{etau}). }
\label{tabU}
\end{table}
\begin{table}[h!]
	\centering
	\begin{tabular}{| c | c | c | c | }
		\hline
		\multicolumn{4}{| c |}{$Z_{\alpha \beta}$} \\
		\hline
		$\beta_{\downarrow}$  & $\alpha=0$ & $\alpha=k$ & $\alpha=(j,m)$ \\ \hline \hline
		$S_F$ & $g$ & $g$ & $g$ \\ \hline
		$S_I$ & $-\frac{\partial \phi_{-1}}{\partial \nu}$ & $1 $ & $0$ \\  \hline
		$S_B$ & $1$ & $1$ & $1$ \\ \hline
		$S_{\sigma}$ & $1$ & $1$ & $1$ \\
		\hline
	\end{tabular}
	\caption{Table of values for the  functional coefficient $Z_{\alpha \beta}$. Here $k=1,\dots,4$,  $\omega=\sqrt{g \kappa \tanh(\kappa h)}$, and derivative to $\nu$ is derivative to the normal of $S_I$ in equilibrium, Eq. (\ref{m11}).}
	\label{tabZ}
\end{table}
\begin{table}[h!]
	\centering
	\begin{tabular}{| c | c | c | c | }
		\hline
		\multicolumn{4}{| c |}{$\Lambda_{\alpha \beta}$} \\
		\hline
		$\beta_{\downarrow}$  & $\alpha=0$ & $\alpha=k$ & $\alpha=(j,m)$ \\ \hline
		\hline
		$S_F$ & $0$ & $0$ & $0$ \\ \hline
		$S_I$ & $0$ & $g_k$ & Right hand side Eq. (\ref{m10}) \\  \hline
		$S_B$ & $0$ & $0$ & $0$ \\ \hline
		$S_{\sigma}$ & $0$ & $0$ & $0$ \\
		\hline
	\end{tabular}
	\caption{Table of values for the  functional coefficient $\Lambda_{\alpha \beta}$. Here $k=1,\dots,4$, and  coefficients $g_k$ are given by the right hand side of Eqs. (\ref{m12}, \ref{m14}, \ref{m19}, \ref{m17}), for $k=1,\dots, 4$, respectively.}
	\label{tabL}
\end{table}
The complete solution for the Laplace equation for the potential, with time-dependent Robin boundary conditions given in Eq. (\ref{m26}) and tables (\ref{tabU}, \ref{tabZ}, \ref{tabL}) is finally obtained using the single layer potential approach \cite{fol,free,green,potential,h,pottheory,dtn,bc2}. We prefer this method to other approaches (Green function method, boundary element method, or double layer potential) because the mixed boundary conditions have constant coefficients, see tables (\ref{tabU}-\ref{tabL}), except the potential for $k=0$, which will be treated in a separate way. We use the Green's representation formula and the fundamental solution of the Laplace problem
\begin{equation}\label{gr}
G(\vec{r},\vec{r}^{\ '})=\frac{1}{4 \pi ||\vec{r}-\vec{r}^{\ '}||}.
\end{equation}
The solution to our problem becomes
\begin{equation}\label{m26b}
\phi (\vec{r},t)=\phi_0+\sum_{k=1}^{4}\xi_k \phi_k +\sum_{j,m} \phi_{jm}
\end{equation}
with partial potential solution for each degree of freedom except $\alpha=0$ given by an integral representation over the whole surface $\Sigma$ of the water
\begin{equation}\label{m27}
\phi_{\alpha}(\vec{r},t)=\sum_{\beta=1}^{4}\iint_{S_{\beta}} \mu_{\alpha \beta} (\vec{r}^{\ '},t) G(\vec{r},\vec{r}^{\ '}) \ dS, \ \ \ \alpha\neq 0, \ \ \beta=1,\dots,4,
\end{equation}
where the multi-index $\alpha=(j,m)$ running over the spectrum of the eigenproblem Eqs. (\ref{e101c}, \ref{m10}) and $\beta$ labels integration over all four sub-surfaces of the closed boundary $\Sigma$.

The source density function $\mu_{\alpha \beta}$ is obtained by solving a Fredholm integral equation of the second kind for each type of degree of freedom $\alpha\neq 0$ and for each sub-surface $\beta$
\begin{equation}\label{m28}
-\frac{1}{2}\mu_{\alpha \beta} (\vec{r},t)+\sum_{\beta'=1}^{4}\iint_{S_{\beta'}}\biggl[ \frac{U_{\alpha \beta'}}{Z_{\alpha \beta'}}G(\vec{r},\vec{r}^{\ '})+\frac{\partial G}{\partial \nu_{\beta'}^{\ '}}(\vec{r},\vec{r}^{\ '})\biggr] \mu_{\alpha \beta'}(\vec{r}^{\ '},t) dS=\frac{\Lambda_{\alpha \beta}}{Z_{\alpha \beta}},
\end{equation}
where, in the same way as above,  the multi-index $\alpha=(j,m)$ running over the spectrum of the eigenproblem Eqs. (\ref{e101c}, \ref{m10}) and $\beta$ labels integration over all four sub-surfaces of the closed boundary $\Sigma$.

For the scattered potential $\phi_0$ we follow a similar approach, except in this case we have homogeneous Robin boundary condition with spatially varying coefficient. Indeed, from table \ref{tabL} we noticed that $\lambda_{0,\beta}=0$ for all surfaces $\beta$. For this reason we can re-write the boundary condition Eq. (\ref{m26}) in the form
\begin{equation}\label{m28b}
\biggl( \frac{U_{0 \beta}}{Z_{0 \beta}}  \phi_{0} +  \frac{\partial \phi_{0}}{\partial \nu_{0 \beta}} \biggr)_{S_{\beta}}=0.
\end{equation}
The scattered potential solution is given by the integral representation over the whole surface $\Sigma$ of the water
\begin{equation}\label{m27c}
\phi_{0}(\vec{r},t)=\sum_{\beta=1}^{4}\iint_{S_{\beta}} \mu_{0 \beta} (\vec{r}^{\ '},t) G(\vec{r},\vec{r}^{\ '}) \ dS,  \ \ \beta=1,\dots,4
\end{equation}
The source density function $\mu_{0 \beta}$ is obtained by solving a homogeneous Fredholm integral equation of the second kind for each type of  sub-surface $\beta$
\begin{equation}\label{m29}
-\frac{1}{2}\mu_{0 \beta} (\vec{r},t)+\sum_{\beta'=1}^{4}\iint_{S_{\beta'}}\biggl[ \frac{U_{0 \beta'}}{Z_{0 \beta'}}G(\vec{r},\vec{r}^{\ '})+\frac{\partial G}{\partial \nu_{\beta'}^{\ '}}(\vec{r},\vec{r}^{\ '})\biggr] \mu_{0 \beta'}(\vec{r}^{\ '},t) dS=0,
\end{equation}
where, similar to the above solutions, the multi-index $\alpha=(j,m)$ running over the spectrum of the eigenproblem Eqs. (\ref{e101c}, \ref{m10}) and $\beta$ labels integration over all four sub-surfaces of the closed boundary $\Sigma$.

In conclusion, the complete solution for the flow potential around the floe Eq. (\ref{m26b}), considering all boundary conditions and all degrees of freedom, is obtained by solving the two sets of Fredholm integral equations of the second kind  Eqs. (\ref{m27}, \ref{m28}) and Eqs. (\ref{m27c}, \ref{m29}) where the Green function is given in Eq. (\ref{gr}) and the coefficients $U, Z, \Lambda$ from the Fredholm integrals are given in tables (\ref{tabU}-\ref{tabL}). This solution includes in a linear combination the incident wave, the scattered waves resulting from the floe considered at rest, the feedback contribution from the rigid motions of the floe, and the feedback contribution from all elastic normal modes excited in the floe.

The rigid motions of the floe (surge, heave, roll and pitch) are described by small oscillations of amplitudes $\xi_k$, $k=1,\dots,4$ at any arbitrary frequency $\omega$, which are obtained by solving  Eqs. (\ref{m22}) in  section \ref{subdybori}. The modifications of the floe shape because of its flexural oscillations  are 
given by the series for $w(r, \theta, t)$ in Eqs. (\ref{m1}), where the time dependent coefficients $b_{jm}(t)$ are given in   Eqs. (\ref{m7}) and Eqs. (\ref{etau}) from section \ref{seccoupl}, and the eigenfunctions are  given by solving the problem in  Eqs. (\ref{e101c}, \ref{m10}). The complete dynamics of the elastic floe is obtained by adding the contribution from the rigid motion and the contributions of all elastic normal modes.

The time evolution of the system is managed as follows. The rigid motions of the floe and the partial potentials associated with the feedback from these motions, $\phi_{k}$ are described by oscillations with an arbitrary frequency $\omega$. This frequency is arbitrary and considered the parameter of the problem. It is related to the wave number $\kappa$ and represents the characteristic  of the incident water wave of potential $\phi_{-1}$. The scattered potential $\phi_{0}$ has the same periodic behavior described by $\omega$. The other partial potentials representing the feedback from elastic modes of the floe are represented by series of the basis of orthogonal normal modes with frequencies $\tau_{jm}$. When the incident wave of frequency $\omega$ interacts with the floe elastic surface, we can project the time dependence  of each normal mode $e^{\pm i \tau_{jm} t}$ on this incoming frequency as a regular complex Fourier series. The response of each boundary mode $\psi_{jm}(r, \theta) a_{jm}(t)$ of $S_I$ to the $\omega$ excitation will be weighted by a Fourier coefficient
$$
\frac{\sin \biggl[ 2 \pi \biggl(\frac{\tau_{jm}}{\omega}-1 \biggr) \biggr]}{2 \pi \biggl(\frac{\tau_{jm}}{\omega}-1 \biggr)}.
$$
These coefficients describe the attenuation of normal modes far from the exciting frequency, as well as the resonance situations. It results that all partial potentials representing flexural modes $\phi_{\alpha}=\phi_{jm}$, and the floe elastic deformation $w$ will have the series terms weighted correspondingly, and acquire  the same oscillating behavior with the frequency $\omega$ as the other quantities in the model. Finally, all arbitrary constants in the partial potentials and floe elastic deformation function $a_{jm}^{0}, b_{jm}^{0}$ are obtained by matching the initial condition of the system. This is possible because of the  orthogonality of the basis of normal modes.

\section{Conclusions}

We have developed a comprehensive mathematical model for the dynamics of floating elastic ice floes under ocean wave excitation that advances beyond the limitations of existing theories. The principal contributions of this work are threefold: (i) simultaneous treatment of both rigid-body motions and elastic flexural deformations with full inertial effects, (ii) extension to arbitrary floe geometries with non-uniform thickness distributions, and (iii) a unified Green function framework that reduces the coupled hydroelastic problem to tractable Fredholm integral equations.

The model treats the floe as a quasi-planar elastic body satisfying a generalized Kirchhoff-Love equation with spatially varying flexural rigidity $D(r,\theta)$, freely floating over an ideal fluid domain governed by Laplace's equation for the velocity potential. By introducing partial potentials for the incident wave, scattered wave, four rigid-body modes (heave, surge, roll, pitch), and the complete spectrum of elastic deformation modes, we decompose the complex fluid-structure interaction into manageable components. The elastic eigenproblem with free-edge boundary conditions provides a complete orthogonal basis of flexural modes, though the non-uniform thickness prevents clean separation into radial and circumferential components except for axisymmetric cases.

The coupling between fluid and structure is captured through kinematic boundary conditions that match normal velocities at the interface and dynamic conditions relating hydrodynamic pressure to elastic response. For rigid motions, we derive explicit expressions using differential geometry to account for the changing orientation of interface normals during oscillations. The elastic modes are characterized by modified natural frequencies that incorporate added mass effects from the surrounding water, expressed through implicit algebraic equations relating the in-vacuum frequencies to fluid-coupled response.

The solution methodology employs single-layer potential representations with the fundamental Green function, reducing all boundary value problems to systems of Fredholm integral equations of the second kind. We systematically organize all boundary conditions on the free surface, interface, seabed, and artificial far-field boundary into compact tensor form with coefficient matrices presented in three tables. This formulation provides a rigorous mathematical foundation for numerical implementation while maintaining analytical transparency.

The model enables investigation of resonance phenomena when incident wave frequencies approach natural vibration modes, wave attenuation through scattering and elastic dissipation, and the relative importance of rigid versus elastic response across different frequency regimes. Extension to include viscous effects through Helmholtz-Hodge decomposition, time-dependent thickness variations from melting, and three-dimensional geometries with significant vertical extent represents promising directions for future research. The framework established here provides essential theoretical groundwork for understanding ice-wave interactions in the marginal ice zone, with direct applications to climate modeling, ice breakup prediction, and wave propagation in polar seas.

\section*{Acknowledgments}

The author gratefully acknowledges E. W. Rogers from the US Naval Research Laboratory, Stennis Space Center, Hancock, MS, for initiating this research problem and for his invaluable scientific guidance throughout this work. His insights and continuous support during our numerous discussions have been instrumental in shaping this investigation. The author also extends sincere thanks to the US Naval Research Laboratory (R. Allard, E. W. Rogers) for sponsoring his participation in the Summer Faculty Research Programs during 2021--2023, which provided the opportunity and resources necessary to develop this comprehensive mathematical framework for ice-wave interactions.

\section{Appendix A}
\label{appA}

In order to provide an example of analytic solution for the eigenproblem Eq. (\ref{e101c}) with free edge boundary conditions Eqs. (\ref{e102}, \ref{e103x}) we assume an axis-symmetric top-down conical shape with radius $R_0$ for the floe with ice thickness $d(r, \theta)=d_1 r+ d_2$ and the condition of slow slope of the cone $d_2\gg d_1 r$. In this case we can develop an asymptotic analysis with expansion parameter $\epsilon =d_1 R_0/d_2$ in the dimensionless variable $\zeta=r/R_0$. We re-write the flexural rigidity Eq. (\ref{ddd}) in the form $D(r,\theta)=D_0 d^3$ and it results
$$
D(r)=D_0 d_2^{3}(1+3 \epsilon \zeta+3\epsilon^2 \zeta^2+\epsilon^3 \zeta^3)
$$
We also expand the eigenfunction and eigenvalue
$$
w=w_0+\epsilon w_1+\epsilon^2 w_2+\dots, \ \ \ \lambda=\lambda_0+\epsilon \lambda_1+\epsilon^2 \lambda_2+\dots
$$
In the leading order $\mathcal{O}(1)$ the equation becomes a biharmonic eigenvalue problem
$$
\triangle^2 w_0=k_0^4 w_0,
$$
where $k_0^4=\lambda_0 \rho_e /(D_0 d_2^2)$. Using regularity at origin and free edge boundary conditions at $r=R_0$ we have the leading term for the eigenfunction expressed as an expansion in Fourier modes $j$
$$
w_0^j(r, \theta)=(A_j J_j (k_0 r)+C_j I_j (k_0 r)) e^{i j \theta}.
$$
The amplitudes $A_j, C_j$ and the spectrum of eigenvalues labeled by $\alpha_{j,m}$ can be obtained from the characteristic equations
$$
\frac{C_j}{A_j}=\frac{J_j (\alpha)}{I_j (\alpha)}, \ \ \ J'_j(\alpha) I_j (\alpha)-J_j (\alpha) I_j'(\alpha)=0
$$
where $\alpha =k_0 R_0$. In the leading order the degenerated spectrum is
$$
\lambda_{0}^{(j, m)}=\frac{D_0 d_2^2 }{\rho_e R_0^4}\alpha_{j,m}^4,
$$ 
where $\alpha_{j,m}$ are the solutions of the second characteristic equations above. The equation for the first order correction $\mathcal{O}(\epsilon)$ is
$$
\triangle^2 w_1-k_0^4 w_1=k_0^4 w_0 \biggl( \frac{\lambda_1}{\lambda_0}-2 \zeta \biggr)
$$
This non-homogeneous biharmonic equation can be solved using the Fredholm alternative condition of solvability. The first order correction in the eigenvalues is 
$$
\lambda_1^{(j,m)}=2 \lambda_0^{(j,m)}\frac{\int_0^{R_0} r^2 R_{j,m}^2 (r) \ dr }{R_0 \int_0^{R_0}r R_{j,m}^2(r) \ dr}
$$

\section{Appendix B}
\label{appB}

The natural frequencies of the elastic mode $(j,m)$ in vacuum, $\omega_{jm}$, occurring in   Eqs. (\ref{m7}) and Eqs. (\ref{etau}) from section \ref{seccoupl}, can be obtained from solving  the eigenproblem in  Eqs. (\ref{e101c}, \ref{m10}). For a general shape this frequencies can be obtained only numerically because 	the free edges boundary conditions $M_{\nu}=0, V_{\nu}=0$ from Eqs. (\ref{e102}, \ref{e103x}) consist in a  highly coupled system, where the thickness variation $d(r,\theta)$ creates strong coupling between radial and azimuthal modes through the gradient terms of $D$, Eq. (\ref{ddd}). Such a procedure can be accomplished in an easier way by using a perturbation approach over  normal modes for an disk of radius $R_0$ and  uniform thickness $d_0$=constant \cite{lei,sode,la,lu,flop,ew1,ew2,ew3,ew4,ew5,ew6}. Such uniform thickness generate of characteristic equation for the frequencies in the form
$$
\left[J_j''(\lambda) + \frac{\nu_e}{\lambda}J_j'(\lambda) + \frac{\nu_e j^2}{\lambda^2}J_j(\lambda)\right]
\left[I_j'''(\lambda) + \frac{1-j^2}{\lambda}I_j''(\lambda) - \frac{1+j^2(1-\nu_e)}{\lambda^2}I_j'(\lambda)\right]
$$
$$
- \left[I_j''(\lambda) + \frac{\nu_e}{\lambda}I_j'(\lambda) + \frac{\nu_e j^2}{\lambda^2}I_j(\lambda)\right]
\left[J_j'''(\lambda) + \frac{1-j^2}{\lambda}J_j''(\lambda) - \frac{1+j^2(1-\nu_e)}{\lambda^2}J_j'(\lambda)\right] = 0,
$$
where $I_j, J_j$ are the corresponding Bessel functions, $j$ is the principal (radial) mode number ($j=0,1,\dots$)
and the solutions for $\lambda$ from the above equation generate the frequencies $\omega_{jm}$ with multiplicity $m=0,1,\dots$
$$
\omega_{jm}=\lambda_{jm}^2\frac{1}{R_0}\sqrt{\frac{E d^2_0}{12 \rho_e (1-\nu^2)}}
$$

\end{document}